\documentclass[twocolumn,aps,prc,10pt,nosuperscriptaddress,nofootinbib,preprintnumbers]{revtex4-1}
\usepackage{amsmath,graphicx,float,csquotes,mdframed,appendix,url}
\usepackage[dvipsnames]{xcolor}
\usepackage[colorlinks=true, pdfstartview=FitV, linkcolor=RedOrange, citecolor=Emerald, urlcolor=Cerulean]{hyperref}
\usepackage{amssymb}
\usepackage{soul}

\newcommand{\eq}[1]{\begin{align} #1 \end{align}}

\newcommand{\Ru}{\rm Ru}
\newcommand{\Zr}{\rm Zr}
\newcommand{\Pb}{\rm Pb}
\newcommand{\U}{\rm U}
\newcommand{\Ox}{\rm O}

\newcommand{\Cu}{\rm Cu}
\newcommand{\Au}{\rm Au}

\begin{document}

\preprint{}
\title{Disentangling baryon stopping and neutron skin effects in heavy-ion collisions}

\author{Gr\'egoire Pihan}
\email{GregoirePihan@uh.edu}
\affiliation{Physics Department, University of Houston, Box 351550, Houston, TX 77204, USA}

\author{Volodymyr Vovchenko}
\email{vvovchen@central.uh.edu}
\affiliation{Physics Department, University of Houston, Box 351550, Houston, TX 77204, USA}

\begin{abstract}
We analyse the net baryon (B) and net electric charge (Q) stopping in heavy-ion collisions using the statistical model. 
Focusing first on isobar collisions $\Ru+\Ru$ and $\Zr+\Zr$ at $\sqrt{s_{\rm NN}}=200$~GeV, we show that the excess baryon-stopping parameter
$\gamma_B \equiv (N_B/N_Q)\,(Z/A)$
can be quantitatively extracted from forthcoming RHIC measurements of charge- and baryon-stopping ratios. 
We then generalize the approach to other collision systems at RHIC and LHC energies and introduce an oxygen-baseline observable, $r^{OX}$, whose central-to-peripheral ratio, $\mathcal{R}^{OX}$, exhibits strong and systematic sensitivity to the neutron-skin thickness of the target nucleus $X$.
\end{abstract}

\maketitle

\section{Introduction}

In ultra-relativistic heavy-ion collisions, the particle yields measured at midrapidity encode essential information on the initial deposition and subsequent transport of conserved charges -- primarily net baryon number ($B$) and net electric charge ($Q$). 
Specifically, the yields of various species of identified charged particles measured at midrapidity are sensitive to the degree of charge stopping, i.e. the longitudinal deceleration of $B$ and $Q$ carried by the incoming protons and neutrons. 

Exploiting this sensitivity, the STAR Collaboration has recently proposed a strategy~\cite{STAR:2024lvy} to probe a baryon-stopping mechanism associated with the baryon junction~\cite{ROSSI1980153}, originally discussed in Ref.~\cite{Kharzeev:1996sq}.
By comparing midrapidity net-baryon and net-charge yields in isobar collisions (Ru+Ru and Zr+Zr) at $\sqrt{s_{\rm NN}}=200$~GeV, one may test whether the transported baryon number remains correlated with valence quarks or instead follows the color-topological baryon junction.

The comparison is quantified by the ratio
\begin{equation}
r = \frac{B}{\Delta Q}\frac{\Delta Z}{A}
\label{eq:ratioSTAR}
\end{equation}
where $\Delta Q = Q_{\mathrm{Ru}} - Q_{\mathrm{Zr}}$ denotes the difference in net electric charge at midrapidity between the two systems,
$B = (B_{\mathrm{Ru}} + B_{\mathrm{Zr}})/2$ is the corresponding average net baryon number at midrapidity,
$A=96$, and $\Delta Z = Z_{\mathrm{Ru}} - Z_{\mathrm{Zr}} = 4$.
Preliminary STAR results~\cite{STAR:2024lvy} indicate $r>1$, suggesting an excess of baryon stopping relative to charge stopping.
Such a hierarchy is consistent with partial decorrelation between baryon and valence-quark transport, as expected in baryon junction-driven stopping scenarios.

However, the interpretation of $r$ is not unique.
Midrapidity yields depend not only on stopping dynamics, but also on (i) the initial proton and neutron distributions in the colliding nuclei---in particular the difference of their root-mean-square radii (the neutron-skin thickness)---and (ii) the subsequent evolution of conserved charges during the medium expansion up to freeze-out.
Consequently, the ratio in Eq.~\eqref{eq:ratioSTAR} reflects a convolution of stopping physics, initial-state nuclear-structure effects, and medium evolution.
Additional complications arise because the observable involves two different nuclei, for which both the nucleon density profiles and the resulting medium evolution can differ.
A quantitative interpretation of measured $r$ values therefore requires a framework that accounts for nuclear-structure effects in a controlled manner.

In this work, we develop such a framework and use it to disentangle the contributions from baryon-junction stopping and neutron-skin physics in ratios of the type~\eqref{eq:ratioSTAR}.
We compute midrapidity hadron yields with the \texttt{Thermal-FIST} package~\cite{Vovchenko:2019pjl}, which evaluates yields in the
hadron resonance gas model at chemical freeze-out for given volume, temperature, and chemical potentials, including resonance-decay feeddown.
We introduce a phenomenological parameter $\gamma_B$ that quantifies excess baryon stopping relative to charge stopping.
By comparing our thermal-model predictions with the hydrodynamic simulations of Ref.~\cite{Pihan:2024lxw}, we calibrate $\gamma_B$ and demonstrate how it can be extracted from experimental measurements of $r$, enabling systematic studies of junction-induced baryon stopping.

We go beyond the isobar case by generalizing Eq.~\eqref{eq:ratioSTAR} to arbitrary pairs of colliding nuclear species, defining ratios $r^{\Gamma X}$ for a reference nucleus $\Gamma$ and a target nucleus $X$.
This generalization allows the construction of observables that primarily isolate the neutron-skin thickness of a single nucleus.
In particular, the centrality dependence of the ratio $r^{OX}$, where $\Ox$ denotes oxygen, provides enhanced sensitivity to the neutron-skin thickness of the nucleus $X$.
Although ${}^{16}\mathrm{O}$ exhibits pronounced $\alpha$-cluster correlations in state-of-the-art nuclear-structure calculations~\cite{Sharma:2025jhs}, for the observables considered here, its density profile is well approximated by a spherical Woods--Saxon distribution with a negligible neutron skin (see Ref.~\cite{Loizides:2025ule}).
Oxygen therefore provides a natural baseline, such that $r^{OX}$ depends predominantly on the neutron-skin properties of $X$.
We evaluate $r^{O X}$ for $X=\mathrm{Cu},\mathrm{Ru},\mathrm{Zr},\mathrm{Au},\mathrm{U}$ (RHIC) and $X=\mathrm{Pb}$  (LHC), and find a strong, systematic dependence on the neutron-skin thickness of the tested nucleus, paving the way for neutron-skin constraints from heavy-ion measurements.

\section{Theoretical framework}
\label{sec:TheoreticalFramework}
\vspace*{-2mm}

\subsection{Model setup}

We study the deposition of net baryon number ($B$) and net electric charge ($Q$) at midrapidity within the statistical hadronization model as implemented in \texttt{Thermal-FIST}~\cite{Vovchenko:2019pjl}. We assume that, at chemical freeze-out, the midrapidity matter can be described as a hadron-resonance gas (HRG) in chemical equilibrium. Hadron yields at midrapidity are obtained from thermal densities of primordial hadrons, supplemented by resonance-decay feeddown. The yields are fully specified by the temperature $T$, the chemical potentials associated with conserved charges ($\mu_B$, $\mu_Q$, $\mu_S$), and the effective volume per unit rapidity, $\mathrm{d}V/\mathrm{d}y$.

We fix $T$ and $\mu_B$ using established chemical freeze-out systematics~\cite{Vovchenko:2015idt,Andronic:2017pug}. Specifically, we consider Pb+Pb ($\sqrt{s_{\rm NN}}=5.02$~TeV) and O+O ($\sqrt{s_{\rm NN}}=5.36$~TeV) collisions at the LHC, as well as Ru+Ru and Zr+Zr ($\sqrt{s_{\rm NN}}=200$~GeV) collisions at RHIC. We set $T=158$~MeV for all systems, while the baryon chemical potential is taken from Refs.~\cite{ALICE:2023ulv,Vovchenko:2015idt}: $\mu_B=0.90$~MeV for Pb+Pb at 5.02~TeV, $\mu_B=0.84$~MeV for O+O at 5.36~TeV, and $\mu_B=21.2$~MeV for Ru+Ru and Zr+Zr at 200~GeV.

The remaining chemical potentials $\mu_Q$ and $\mu_S$ are determined by imposing a fixed charge-to-baryon ratio at midrapidity and strangeness neutrality, i.e.\ by solving
\begin{align}
    \frac{n^X_Q}{n^X_B} &= \frac{p^X_{\rm frac}(c)}{\gamma_B},
    \label{eq:Qcond}\\
    n^X_S &= 0.
    \label{eq:Scond}
\end{align}
Here $n_B$, $n_Q$, and $n_S$ denote the net-baryon, net-electric-charge, and net-strangeness densities evaluated in the HRG model, $c$ labels the centrality class, and $X$ denotes the collision system. The strangeness-neutrality condition in Eq.~\eqref{eq:Scond} is a standard assumption for midrapidity at RHIC and LHC energies, although scenarios with non-vanishing local strangeness have been discussed (see Ref.~\cite{Ross:2025qxr}). The charge-to-baryon constraint \eqref{eq:Qcond}, on the other hand, is a central element of our formalism: it generalizes the commonly used global constraint $n_Q/n_B=Z/A$ by incorporating (i) an \emph{excess baryon stopping} parameter $\gamma_B$ \cite{Grefa:2026meq} and (ii) the \emph{participant proton fraction} $p^X_{\rm frac}(c)$.
For what follows, the two ingredients entering Eq.~\eqref{eq:Qcond} play qualitatively different roles: $p_{\rm frac}^X(c)$ primarily controls the \emph{centrality dependence} of the observables through nuclear geometry and neutron-skin effects, whereas $\gamma_B$ mainly acts as an approximately centrality-independent \emph{normalization} that shifts the overall magnitude of the stopping ratios.

The parameter $\gamma_B$ quantifies the excess amount of baryon stopping relative to electric-charge stopping at midrapidity. For $\gamma_B=1$, baryon and charge stopping are identical, as expected if both $B$ and $Q$ are transported predominantly by valence quarks. Values $\gamma_B>1$ imply additional baryon transport beyond what is implied by electric charge, indicating a partial dynamical decorrelation of baryon and charge carriers, consistent with an additional mechanism such as baryon-junction transport.

The participant proton fraction is defined as
\begin{equation}
    p^X_{\rm frac}(c) =
    \frac{\langle p^X\rangle_{\rm part}(c)}{\langle p^X+n^X\rangle_{\rm part}(c)},
    \label{eq:pfrac}
\end{equation}
where $\langle p^X\rangle_{\rm part}(c)$ and $\langle p^X+n^X\rangle_{\rm part}(c)$ are the event-averaged numbers of participant protons and participant nucleons, respectively, in centrality class $c$ of $X+X$ collisions. In the presence of different spatial distributions of protons and neutrons in nucleus $X$, $p^X_{\rm frac}(c)$ deviates from the global charge fraction $Z/A$ and becomes centrality dependent.

We estimate $p^X_{\rm frac}(c)$ using the 3D Glauber model of Ref.~\cite{Shen:2020jwv}. In this model, the proton and neutron density profiles, $\rho_p$ and $\rho_n$, are described by Woods--Saxon distributions,
\begin{equation}
    \rho_{p,n}^X(r,\theta,\varphi)=
    \frac{\rho_0^X}{1+\exp\!\left(\frac{r-R_{p,n}^{X,\rm WS}(\theta,\varphi)}{a_{p,n}^X}\right)}\,,
    \label{eq:WSprofile1}
\end{equation}
where $a_{p,n}^X$ are the diffuseness parameters and $R_{p,n}^{X,\rm WS}$ are the (possibly deformed) radius parameters. We parametrize the angular dependence of $R_{p,n}^{X,\rm WS}$ as
\begin{align}
    R_{p,n}^{X,\rm WS}(\theta,\varphi)
    & =
    R_{p,n}^X\!\left[1+\beta_2 Y_2^0(\theta,\varphi)+\beta_3 Y_3^0(\theta,\varphi) \right. \nonumber \\
    & \qquad \left. + \beta_4 Y_4^0(\theta,\varphi)\right],
\end{align}
where $Y_\ell^0$ are spherical harmonics, $R_{p,n}^X$ are the mean half-density radii, and $\beta_{2,3,4}$ are the quadrupole, octupole, and hexadecapole deformation parameters, respectively.

Table~\ref{table:WSParams} summarizes the Woods--Saxon parameters used for the nuclei considered in this work. The proton half-density radius $R_p$, proton diffuseness $a_p$, and deformation parameters $\beta_\ell$ are taken from the literature. We assume that neutron-skin effects are primarily driven by a difference in diffuseness, $\Delta a \equiv a_n-a_p$, and take the difference between proton and neutron half-density radii to be negligible. Specifically, we set $R_n-R_p=0.01$~fm, consistent with the convention adopted in Ref.~\cite{Giacalone:2023cet}. The neutron diffuseness $a_n$ is then fixed to reproduce literature values of the neutron-skin thickness $\Delta R_{\rm np}$ (listed in Table~\ref{table:WSParams}), defined as the difference between the root-mean-square (RMS) radii of neutron and proton density distributions. To relate $\Delta R_{\rm np}$ to $\Delta a$, we compute RMS radii from the Woods--Saxon profiles in Eq.~\eqref{eq:WSprofile1}.

The neutron-skin thickness of ${}^{63}{\rm Cu}$ is not well constrained. We therefore estimate it using a linear fit (with zero intercept, following Ref.~\cite{Zhang:2021jwh}) to literature values of $\Delta R_{\rm np}$ as a function of $(N-Z)/A$. This yields $\Delta R_{\rm np}\simeq 1.02\,(N-Z)/A$, implying $\Delta R_{\rm np}\simeq 0.08$~fm for ${}^{63}{\rm Cu}$.

We take ${}^{16}{\rm O}$ as the baseline nucleus in our calculations. For oxygen we neglect deformations ($\beta_{2,3,4}=0$) and assume identical proton and neutron distributions, $R_p=R_n$ and $a_p=a_n$, such that $\Delta R_{\rm np}=0$. We thus ignore known $\alpha$-cluster correlations in ${}^{16}{\rm O}$~\cite{Behera:2021zhi,Sharma:2025jhs} and leave their possible impact on the present observables for future work.

\begin{figure}
    \centering
    \includegraphics[width=0.46\textwidth]{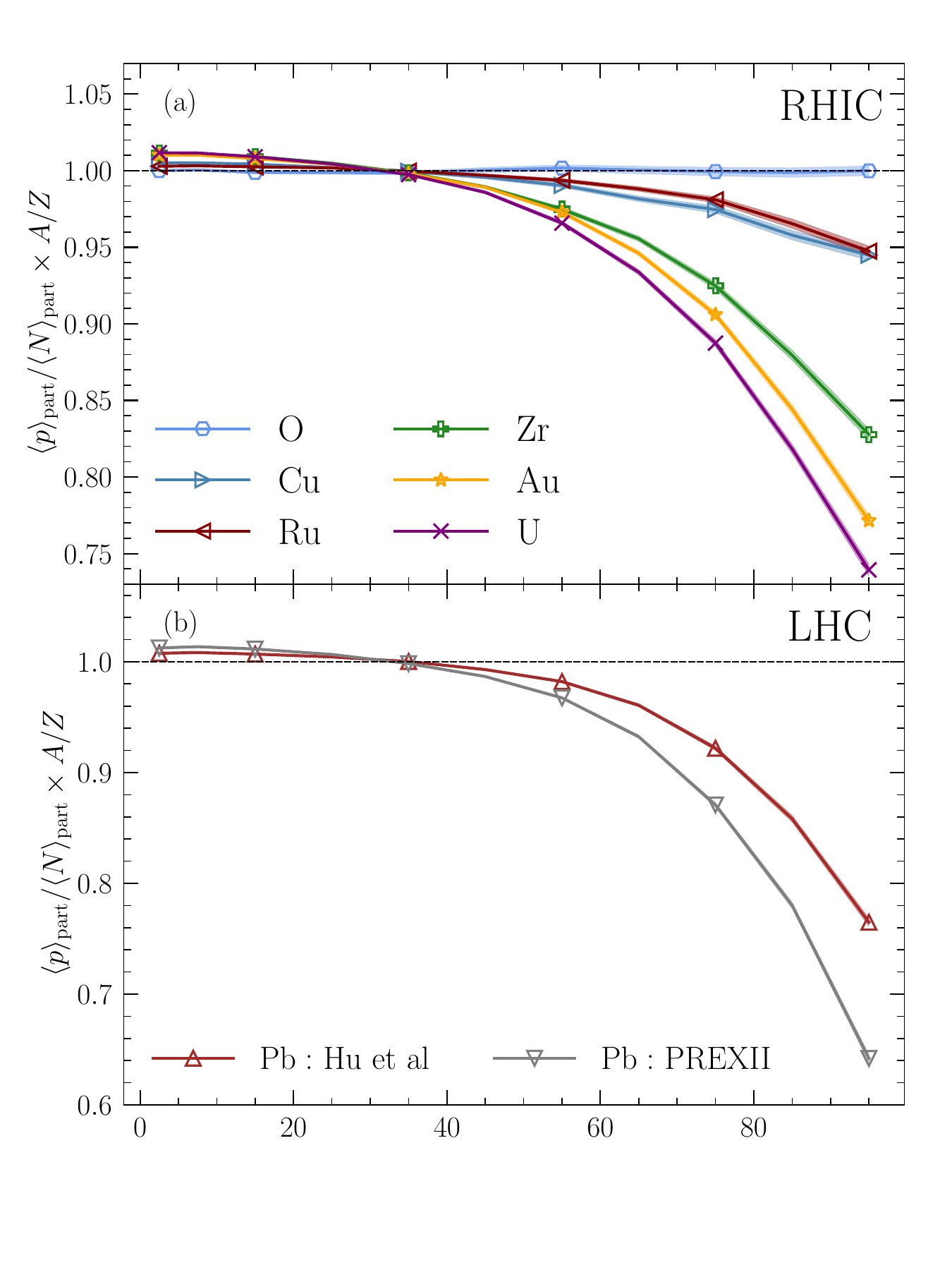}
    \caption{Participant proton fraction $p_{\rm frac}(c)$ obtained from the 3D Glauber model using the Woods--Saxon parameters in Table~\ref{table:WSParams}, shown as a function of centrality and normalized by the global charge fraction $Z/A$. Panel (a): $p_{\rm frac}(c)/(Z/A)$ for $\Ox$, $\Cu$, $\Ru$, $\Zr$, $\Au$ and $\U$. 
    Panel (b): the same quantity for Pb for two choices of neutron-skin thickness, $\Delta R_{\rm np}\simeq 0.17$~fm from \emph{ab initio} calculations~\cite{Hu:2021trw} and $\Delta R_{\rm np}\simeq 0.28$~fm from PREX-II~\cite{PREX:2021umo}.
    }
    \label{fig:pfracGlauber}
\end{figure}

Figure~\ref{fig:pfracGlauber} shows the participant proton fraction for O, Cu, Zr, Ru, Au, U, and Pb as a function of centrality, normalized to the global charge fraction $Z/A$. Panel (a) reveals a pronounced centrality dependence for Zr, Au, and U, reflecting the depletion of participant protons in peripheral collisions due to neutron-skin effects. Panel (b) compares $p_{\rm frac}^{\rm Pb}(c)$ for two values of the Pb neutron-skin thickness: an \emph{ab initio} estimate $\Delta R_{\rm np}\simeq 0.17$~fm~\cite{Hu:2021trw} and the larger value inferred by PREX-II, $\Delta R_{\rm np}\simeq 0.28$~fm~\cite{PREX:2021umo}. Increasing $\Delta R_{\rm np}$ enhances the depletion of participant protons in peripheral events and, by overall charge conservation, leads to a compensating enhancement in central collisions.

The final ingredient is the effective system volume at midrapidity, $\mathrm{d}V/\mathrm{d}y$. 
Because we compare different collision systems and centralities, the treatment of $\mathrm{d}V/\mathrm{d}y$ requires particular care.
We assume that the freeze-out volume per unit rapidity scales with the charged-particle multiplicity at midrapidity, $\mathrm{d}N_{\rm ch}^X/\mathrm{d}\eta|_{\eta=0}$, such that the \emph{relative} volume between systems $X$ and $Y$ in the same centrality bin is
\begin{equation}
    \frac{\mathrm{d}V^X/\mathrm{d}y}{\mathrm{d}V^Y/\mathrm{d}y}
    \simeq
    \frac{\left.\mathrm{d}N_{\rm ch}^X/\mathrm{d}\eta\right|_{\eta=0}(c)}
         {\left.\mathrm{d}N_{\rm ch}^Y/\mathrm{d}\eta\right|_{\eta=0}(c)}.
    \label{eq:volumeRatio}
\end{equation}

The charged multiplicity at midrapidity is estimated using the standard Glauber-based two-component model~\cite{Kharzeev:2000ph},
\begin{equation}
    \left.\frac{\mathrm{d}N_{\rm ch}^X}{\mathrm{d}\eta}\right|_{\eta=0}(c)
    \simeq
    n_{pp}\left[
    \frac{1-x}{2}\,N_{\rm part}(c)
    + x\,N_{\rm coll}(c)
    \right],
    \label{eq:dNch}
\end{equation}
where $n_{pp}$ is the charged multiplicity at midrapidity in $p{+}p$ collisions at the same energy, and $x$ denotes the hard-scattering fraction. 
We take $x=0.12$ for both RHIC and LHC energies throughout~\cite{Zhou:2013zsa}.

To set the overall scale, we use the volume in the $0$--$5\%$ most central Au+Au collisions at RHIC as a reference,
\begin{equation}
    \left.\frac{\mathrm{d}V^{\rm Au}}{\mathrm{d}y}\right|_{0\text{--}5\%}
    = 1000~\mathrm{fm}^3,
\end{equation}
where the precise numerical value is unimportant for the ratios considered here, as it cancels in the final observables.\footnote{Only relative volumes between systems and centralities enter the ratios studied in this work.}
The volume for any system $X$ and centrality class $c$ is then obtained via
\begin{equation}
    \left.\frac{\mathrm{d}V^X}{\mathrm{d}y}\right|(c)
    =
    \frac{\left.\mathrm{d}V^{\rm Au}/\mathrm{d}y\right|_{0\text{--}5\%}}
         {\left.\mathrm{d}N_{\rm ch}^{\rm Au}/\mathrm{d}\eta\right|_{\eta=0,\,0\text{--}5\%}}
    \times
    \left.\frac{\mathrm{d}N_{\rm ch}^X}{\mathrm{d}\eta}\right|_{\eta=0}(c).
    \label{eq:dV}
\end{equation}
Note that $n_{pp}$ is independent of centrality and therefore cancels in Eq.~\eqref{eq:dV}. 
The quantities $N_{\rm part}$ and $N_{\rm coll}$ fluctuate event-by-event, which would lead to a fluctuating volume.
We obtain the average volume by simulating 100k 3D Glauber model events.

With all HRG inputs fixed, we compute midrapidity hadron yields using \texttt{Thermal-FIST}~\cite{Vovchenko:2019pjl}. 
We employ the standard PDG2020 particle list within the ideal HRG framework and neglect finite resonance widths. 
Decay feeddown is included, with weak decays treated as discussed below.

\begin{table*}
\centering
\caption{Woods--Saxon (WS) parameters used for the nuclei considered in this study. Unless stated otherwise, we take $R_n=R_p+0.01$~fm;  
Neutron-skin effects are modeled primarily through a difference in diffuseness, $\Delta a \equiv a_n-a_p$, while keeping the difference in half-density radii negligible. For each nucleus, the neutron diffuseness $a_n$ is determined by matching the RMS neutron-skin thickness $\Delta R_{\rm np}$ (last column), defined as the difference between the neutron and proton RMS radii computed from Eq.~\eqref{eq:WSprofile1}. 
(a) ${}^{16}\mathrm{O}$ is treated as spherical and undeformed ($\beta_{2,3,4}=0$) with $\Delta R_{\rm np}=0$. 
(b) For ${}^{63}\mathrm{Cu}$ we estimate $\Delta R_{\rm np}$ from a linear fit of literature values versus $(N-Z)/A$ following Ref.~\cite{Zhang:2021jwh}. 
For ${}^{208}\mathrm{Pb}$ we consider two benchmark values of $\Delta R_{\rm np}$, $\simeq 0.17$~fm from \emph{ab initio} calculations~\cite{Hu:2021trw} and $\simeq 0.28$~fm from PREX-II~\cite{PREX:2021umo}.}
\begin{tabular}{|c | c | c | c | c | c | c | c | c | c |} 
 \hline
  & $R_\mathrm{p}$ (fm)&  $R_\mathrm{n}$ (fm) & $a_\mathrm{p}$ (fm) & $a_\mathrm{n}$ (fm) & $\beta_2$ & $\beta_3$ & $\beta_4$ & $\Delta R_{\mathrm{np}}$ (fm) \\ 
  \hline
  ${}^{16}\Ox$ & 2.608 \cite{DeVries:1987atn} &  2.608 & 0.513 \cite{DeVries:1987atn} & 0.513 & $0^{(a)}$ & $0^{(a)}$ & $0^{(a)}$ & $0^{(a)}$ \\ 
 \hline
 ${}^{96}\Ru$ & 5.09 \cite{Xu:2021vpn} & 5.1 & 0.46 \cite{Schenke:2020mbo} & 0.488 & 0.16 \cite{Zhang:2021kxj} & 0 \cite{Moller:2015fba} & 0 \cite{Moller:2015fba} & 0.05 \cite{Liu:2023qeq}\\ 
 \hline
  ${}^{63}\Cu$  & 4.2 \cite{Miller:2007ri} & 4.21 & 0.598 \cite{Miller:2007ri} & 0.632 & 0.151 \cite{Moller:2015fba} & 0 \cite{Moller:2015fba} & -0.016 \cite{Moller:2015fba} & $0.08^{(b)}$\\ 
 \hline
 ${}^{96}\Zr$ & 5.02  \cite{Xu:2021vpn} & 5.03 & 0.52 \cite{Zhang:2021kxj} & 0.618 & 0.06 \cite{Zhang:2021kxj} & 0.2 \cite{Zhang:2021kxj} & 0 \cite{Schenke:2020mbo} & 0.18 \cite{Liu:2023qeq} \\ 
 \hline
  ${}^{197}\Au$ & 6.38 \cite{Miller:2007ri} & 6.39 & 0.535 \cite{Miller:2007ri} & 0.662 & -0.13 \cite{Schenke:2020mbo} & 0 \cite{Moller:2015fba} & -0.03 \cite{Schenke:2020mbo} & 0.2 \cite{Liu:2023qeq} \\ 
  \hline
  ${}^{208}\Pb$ & 6.68 \cite{Giacalone:2023cet} & 6.69 & 0.448 \cite{Giacalone:2023cet} & 0.575/0.650 & 0 \cite{Schenke:2020mbo} & 0 \cite{Moller:2015fba} & 0 \cite{Schenke:2020mbo}& 0.17/0.28 \cite{Hu:2021trw}/\cite{PREX:2021umo}\\ 
  \hline
  ${}^{238}\U$ & 6.8 \cite{DeVries:1987atn} & 6.81 & 0.61 \cite{DeVries:1987atn} & 0.759 & 0.25 \cite{Ryssens:2023fkv} & 0 \cite{Moller:2015fba} & 0.124 \cite{Bemis:1973zza} & 0.24 \cite{Liu:2023qeq}
  \\ 
  \hline
\end{tabular}
\label{table:WSParams}
\end{table*}

\subsection{Observables}

We introduce and analyze a generalized version of the observable $r$ in Eq.~\eqref{eq:ratioSTAR}, originally proposed for isobar collisions. Here we extend the construction to arbitrary pairs of colliding nuclei.

We start from the global charge conservation (GCC) expectation for symmetric $X{+}X$ collisions. In the GCC scenario, the charge-to-baryon ratio at midrapidity satisfies
$Q^X/B^X = Z_X/A_X$,
where $Z_X$ and $A_X$ are the proton and mass numbers of nucleus $X$. To quantify deviations from GCC, we define
\begin{equation}
    \mathcal{F}_X
    \equiv
    \frac{Q^X}{\sigma^X}
    - \frac{Z_X}{A_X}\frac{B^X}{\sigma^X}
    =
    \tilde Q^X - \frac{Z_X}{A_X}\tilde B^X,
    \label{eq:sigma}
\end{equation}
where $\sigma^X$ is an optional scale parameter associated with system $X$ that facilitates comparisons between collision systems of different sizes, and we introduced the reduced quantities $\tilde Q^X \equiv Q^X/\sigma^X$ and $\tilde B^X \equiv B^X/\sigma^X$. In practice, we consider two choices: no scaling, $\sigma^X=1$, and multiplicity scaling, $\sigma^X=\langle N_{\rm ch}^X\rangle_{\rm ev}$, where $\langle N_{\rm ch}^X\rangle_{\rm ev}$ is the mean charged-particle multiplicity at midrapidity in the corresponding centrality class.

Independent of the choice of $\sigma^X$, the GCC condition implies
\begin{equation}
    \mathcal{F}_X = 0.
    \label{eq:GCC}
\end{equation}
Within our framework, $\mathcal{F}_X$ can be expressed in terms of the excess baryon-stopping parameter $\gamma_B$ and the participant proton fraction $p_{\rm frac}^X(c)$, and thus in principle provides sensitivity to baryon-junction transport and neutron-skin effects.

In practice, however, $\mathcal{F}_X$ requires an accurate determination of the net charge $Q^X$ at midrapidity. Since $Q^X$ is typically much smaller than the individual yields of positively and negatively charged hadrons, its direct measurement often carries sizeable uncertainties. This limitation can be mitigated by considering \emph{differences} between systems, for which relative changes in net charge can be determined with significantly improved precision, as demonstrated by the STAR isobar analysis~\cite{STAR:2024lvy}.

We therefore define the difference operator
\begin{equation}
    \Delta \mathcal{O} \equiv \mathcal{O}_X - \mathcal{O}_Y,
\end{equation}
which denotes the difference of an observable $\mathcal{O}$ between $X{+}X$ and $Y{+}Y$ collisions. Applied to $\mathcal{F}$, this yields
\begin{equation}
    \Delta \mathcal{F}
    =
    \mathcal{F}_X-\mathcal{F}_Y
    =
    \Delta \tilde Q
    - \frac{Z_X}{A_X}\tilde B^X
    + \frac{Z_Y}{A_Y}\tilde B^Y,
\end{equation}
where $\Delta \tilde Q \equiv \tilde Q^X - \tilde Q^Y$.
If GCC holds for both systems, Eq.~\eqref{eq:GCC} implies $\Delta\mathcal{F}=0$. The converse also holds under two mild assumptions that we adopt throughout: (i) there is no additional mechanism that generates net electric charge at midrapidity beyond that implied by the transported baryon number, and 
(ii) $\sigma^X$ is chosen independently of isospin (here we use $\sigma^X=1$ or $\sigma^X=\langle N_{\rm ch}^X\rangle_{\rm ev}$), rather than a normalization tied to $(Z_X/A_X)N_{\rm part}^X$ which can partially absorb the leading system-size dependence.

Provided that $\Delta\tilde Q\neq 0$, the condition $\Delta\mathcal{F}=0$ can be rewritten as
\begin{equation}
    \frac{Z_X}{A_X}\frac{\tilde B^X}{\Delta\tilde Q}
    -
    \frac{Z_Y}{A_Y}\frac{\tilde B^Y}{\Delta\tilde Q}
    = 1.
\end{equation}
We define the left-hand side as the generalized ratio $r^{XY}$, such that the GCC expectation takes the compact form
\begin{equation}
    r^{XY}=1.
    \label{eq:GCCratio}
\end{equation}
After straightforward algebra, $r^{XY}$ can be expressed as
\begin{equation}
    r^{XY}
    =
    \alpha_Y\,\frac{\Delta \tilde B}{\Delta \tilde Q}
    + \Gamma_{XY}\,\frac{\tilde B^X}{\Delta \tilde Q},
    \label{eq:ratio}
\end{equation}
where $\Delta \tilde B \equiv \tilde B^X-\tilde B^Y$, and we introduced the coefficients
\eq{
\alpha_Y = \frac{Z_Y}{A_Y}, \qquad
\Gamma_{XY} = \frac{\Delta Z}{A_X} - \frac{Z_Y\,\Delta A}{A_X A_Y},
}
with $\Delta Z \equiv Z_X-Z_Y$ and $\Delta A \equiv A_X-A_Y$.

Note that Eq.~\eqref{eq:ratio} involves only the difference $\Delta \tilde Q$ of the (normalized) midrapidity net charge, rather than the individual charges $\tilde Q^X$ and $\tilde Q^Y$. This is advantageous experimentally, because $\Delta \tilde Q$ can be constructed in a way that reduces correlated systematic effects when comparing two systems. While such cancellations are most effective for isobars (where data can be taken under nearly identical detector conditions), the same strategy remains useful more generally, especially when the scaling choices are such that residual correction terms are small.

Following Ref.~\cite{STAR:2024lvy}, we construct $\Delta \tilde Q$ from double ratios of identified-hadron yields at midrapidity,
\begin{equation}
    R2^{XY}_{\tilde{h}} =
    \frac{\tilde{h}^X_+/\tilde{h}^X_-}{\tilde{h}^Y_+/\tilde{h}^Y_-},
    \label{eq:R2}
\end{equation}
where $\tilde h^X_{\pm}\equiv h^X_{\pm}/\sigma^X$ denote the yields of positively/negatively charged hadrons
$h\in\{\pi,K,p\}$ at midrapidity in $X{+}X$ collisions, normalized by the scale factor $\sigma^X$ introduced above.
As shown in Appendix~\ref{sec:AppA}, to first order in the charge-asymmetry and system-mismatch parameters one can express the difference in net-$h$ number between the two systems as
\begin{align}
    \Delta \tilde h
    &\equiv
    \left(\tilde h^X_{+}-\tilde h^X_{-}\right)
    -
    \left(\tilde h^Y_{+}-\tilde h^Y_{-}\right)
    \nonumber\\
    &=
    (1+\tilde{\beta}^{XY})\,(1-\tilde{\epsilon}^{Y})\,\hat h^{Y}\,
    \left(R2^{XY}_{\tilde{h}}-1\right),
    \label{eq:DeltaPhi}
\end{align}
with
\begin{align}
    \tilde{\beta}^{XY} &\equiv
    \frac{\tilde{h}^X_{-}-\tilde{h}^Y_{-}}{\tilde{h}^Y_{-}},
    \label{eq:beta}\\
    \tilde{\epsilon}^{Y} &\equiv
    \frac{\hat h^{Y}-\tilde h^{Y}_{-}}{\hat h^{Y}},
    \label{eq:epsilon}
\end{align}
and $\hat h^{Y}\equiv(\tilde h^Y_{+}+\tilde h^Y_{-})/2$ the charge-averaged yield of hadron species $h$ in system $Y$.

The parameter $\tilde{\beta}^{XY}$ quantifies the relative mismatch of the negatively charged yield between systems $X$ and $Y$ and explicitly depends on the scaling choices $\sigma^X$ and $\sigma^Y$. The parameter $\tilde{\epsilon}^{Y}$ quantifies the intrinsic charge asymmetry in system $Y$ (i.e.\ the deviation between $\tilde h_-^Y$ and the charge-averaged yield) and is independent of the overall normalization. Both effects break a perfect one-to-one correspondence between $\Delta \tilde h$ and the double ratio $R2^{XY}_{\tilde h}$. In the setups considered here, however, their numerical values are small (see Tabs~\ref{table:BetaXY} and \ref{table:epsilonY} in Appendix \ref{sec:AppA}), so that the double-ratio construction provides an accurate approximation for $\Delta \tilde h$.

The difference in normalized net electric charge is then obtained by summing over hadron species,
\begin{equation}
    \Delta \tilde Q = \sum_{h\in\{\pi,K,p\}} \Delta \tilde h.
    \label{eq:DeltaQ}
\end{equation}

The net baryon number entering Eq.~\eqref{eq:ratio} is constructed, following Ref.~\cite{STAR:2024lvy}, as the sum of final-state protons and neutrons,
\begin{equation}
    \tilde{B}^{X} =
    \frac{1}{\sigma^X}\Big[
    N_p^X - N_{\bar p}^X + N_n^X - N_{\bar n}^X
    \Big],
    \label{eq:NBN}
\end{equation}
where $N_{p,\bar p}$ ($N_{n,\bar n}$) denote the midrapidity yields of (anti)protons and (anti)neutrons, respectively.
While $\tilde B^X$ can be evaluated directly in the statistical model, experiments do not measure (anti)neutron yields. In Ref.~\cite{STAR:2024lvy} they are reconstructed from other measurements, and we summarize that procedure here.

We separate proton and neutron yields into primordial and feeddown components (dropping the system index $X$ for clarity),
\begin{align}
    N_p &= N_p^{\rm prim} + N_p^{\rm FD}, \qquad
    N_{\bar p} = N_{\bar p}^{\rm prim} + N_{\bar p}^{\rm FD}, \\
    \label{eq:Nn}
     N_n &= N_n^{\rm prim} + N_n^{\rm FD}, \qquad
    N_{\bar n} = N_{\bar n}^{\rm prim} + N_{\bar n}^{\rm FD}.
\end{align}
The primordial (anti)neutron yields are estimated using primordial (anti)protons and measured (anti)deuterons through a statistical-model motivated relation in the Boltzmann approximation,
\begin{align}
\label{eq:Nnprim}
    N_n^{\rm prim} &= N_{\bar p}^{\rm prim}\sqrt{\frac{N_d}{N_{\bar d}}}, \\
\label{eq:Nnbarprim}
    N_{\bar n}^{\rm prim} &= N_{p}^{\rm prim}\sqrt{\frac{N_{\bar d}}{N_d}} .
\end{align}
Here $N_{p,\bar p}^{\rm prim}$ are primordial proton and antiproton yields (i.e.\ before weak-decay feeddown but including strong and electromagnetic decays). Finally, $N_{d,\bar d}$ are the measured deuteron and antideuteron yields.
The weak-decay feeddown contributions to proton and neutron yields are estimated from hyperon yields and the corresponding decay branching ratios (PDG values)
\begin{align}
    N_p^{\rm FD} &= 0.639\!\left(N_{\Lambda}+N_{\Xi^{0}}+N_{\Xi^{-}}+N_{\Omega}\right)\nonumber \\
    & \qquad  + 0.5157\,N_{\Sigma^{+}} ,
    \label{eq:NpFD}\\
    N_n^{\rm FD} &= 0.361\!\left(N_{\Lambda}+N_{\Xi^{0}}+N_{\Xi^{-}}+N_{\Omega}\right) \nonumber \\
    & \qquad + 0.4831\,N_{\Sigma^{+}} + N_{\Sigma^{-}}.
    \label{eq:NnFD}
\end{align}
The feeddown contributions to the yields of antiprotons and antineutrons are estimated through Eqs.~\eqref{eq:NpFD} and \eqref{eq:NnFD} using the yields of antihyperons.
The primordial proton yields $N_p^{\rm prim}$ are computed as $N_p^{\rm prim} = N_p - N_p^{\rm FD}$, primordial neutron yields through Eqs.~\eqref{eq:Nnprim} and \eqref{eq:Nnbarprim}, and the final neutron yields through Eq.~\eqref{eq:Nn}.

The absolute particle yields in $X{+}X$ and $Y{+}Y$ collisions can differ substantially when the colliding nuclei have very different mass numbers. In such cases, the raw yield differences entering $r^{XY}$ may be dominated by trivial system-size effects rather than by the physics of interest. There are two practical strategies to mitigate this issue.

First, one may choose \emph{different} centrality selections in $X{+}X$ and $Y{+}Y$ such that either $\Delta B \simeq 0$ or the midrapidity charged multiplicities are approximately matched, $\left.\mathrm{d}N_{\rm ch}^X/\mathrm{d}\eta\right|_{\eta=0} \simeq \left.\mathrm{d}N_{\rm ch}^Y/\mathrm{d}\eta\right|_{\eta=0}$, for the bins being compared. In this case no explicit normalization is required, and one can set $\sigma^X=\sigma^Y=1$. This approach is feasible when the systems are similar in size (small $\Delta A$ and $\Delta Z$), as is the case for isobar collisions. 

For more disparate nuclei, matching centrality bins in this way may be impractical. A second, more general approach is therefore to compare the same centrality percentiles while introducing a scaling choice $\sigma^X$ and $\sigma^Y$ that reflects the overall particle-production scale in each system (e.g.\ an effective freeze-out volume). In this work we adopt multiplicity scaling,
\begin{equation}
    \sigma^X=\langle N_{\rm ch}^X\rangle_{\rm ev}, \qquad
    \sigma^Y=\langle N_{\rm ch}^Y\rangle_{\rm ev},
\end{equation}
with $\langle N_{\rm ch}^{X,Y}\rangle_{\rm ev}$ estimated from the 3D Glauber model as described earlier in Sec.~\ref{sec:TheoreticalFramework}.

As a validation of the experimental proxy procedure used to reconstruct conserved charges [Eqs.~\eqref{eq:NBN} and~\eqref{eq:DeltaQ}], Table~\ref{tab:proxy_validation} compares the resulting $\tilde B$ and $\Delta\tilde Q$ to the corresponding quantities computed directly in \texttt{Thermal-FIST}. 
We show results for (i) a ``density-matched'' implementation, where we enforce $n_B^{X}=n_B^{Y}$ by adjusting $\mu_B$ between systems (only for isobar case $r^{\Zr \Ru}$), and (ii) the multiplicity-scaled implementation (for all $r^{OX}$ ratios). 
The numerical values correspond to the largest difference observed within all nuclei considered in this study. We also indicate the impact of including the small correction factors in Eq.~\eqref{eq:DeltaPhi} through $\tilde\beta^{XY}$ and $\tilde\epsilon^Y$.
\begin{table*}
\centering
\caption{Maximum relative differences (in \%) between the conserved-charge proxies used in the STAR strategy [Eqs.~\eqref{eq:NBN} and~\eqref{eq:DeltaQ}] and the corresponding values computed directly in \texttt{Thermal-FIST}. The ``fixed-volume'' case (applicable to isobars) corresponds to fixing the volume and net-baryon density of Ru using Zr as a reference. The ``scaled'' case corresponds to $\sigma^X=\langle N_{\rm ch}^X\rangle_{\rm ev}$ and $\sigma^Y=\langle N_{\rm ch}^Y\rangle_{\rm ev}$. ``Corrections'' denotes inclusion of the factors defined in Eqs.~\eqref{eq:beta} and~\eqref{eq:epsilon}; in the first two columns these are neglected.
}
\begin{tabular}{|c | c | c | c | c |}  
 \hline
   & Density-matched & Multiplicity-scaled & Density-matched + corrections & Multiplicity-scaled + corrections \\
  \hline
  $\max|\delta \tilde{B}/\tilde{B}|$ & 0.3\% & 0.3\% & 0.3\% & 0.3\% \\
  \hline
  $\max|\delta \Delta \tilde{Q} / \Delta \tilde{Q}|$ &  1.58\% & 2.2\% & 1.30\% & 2.0\%\\
  \hline
\end{tabular}
\label{tab:proxy_validation}
\end{table*}

Overall, the proxy-based reconstruction reproduces the conserved-charge quantities at the few-percent level, indicating that the STAR procedure provides a robust estimator of $\tilde B$ and $\Delta\tilde Q$ for the RHIC isobar conditions considered here. The fixed-volume and multiplicity-scaled implementations yield comparable accuracy, with a modest improvement in the fixed-volume setup. Including the small correction factors in Eq.~\eqref{eq:DeltaPhi} leads to a further (albeit minor) improvement in the reconstruction of $\Delta\tilde Q$.

\section{Results}

\subsection{Isobar collisions}

We begin with isobar collisions, taking $X=\Zr$ and $Y=\Ru$. Since $A_{\Ru}=A_{\Zr}\equiv A$, we have $\Delta A=0$, and the two nuclei also have similar proton numbers, $Z_{\Ru}=44$ and $Z_{\Zr}=40$. This proximity allows us to explore two prescriptions for matching particle-production scales, introduced in Sec.~\ref{sec:TheoreticalFramework}, and thereby assess the robustness of the extracted baryon-stopping parameter.

\paragraph*{(i) Density-matched setup ($\sigma^{\Ru}=\sigma^{\Zr}=1$).}
Exploiting the similarity of the isobars, we set $\sigma^{\Ru}=\sigma^{\Zr}=1$ and assume equal freeze-out volume per unit rapidity,
$\mathrm{d}V^{\Ru}/\mathrm{d}y=\mathrm{d}V^{\Zr}/\mathrm{d}y$.
We further enforce equal midrapidity net-baryon content,
$\hat B \equiv B^{\Ru}=B^{\Zr}=(B^{\Ru}+B^{\Zr})/2$,
by matching the net-baryon density in the HRG model. Concretely, we compute $n_B^{\Zr}$ at fixed $\mu_B$, and then adjust the value of $\mu_B$ for Ru+Ru such that
\begin{equation}
    n_B^{\Ru}=n_B^{\Zr}.
\end{equation}
In this setup, Eq.~\eqref{eq:ratio} reduces to
\begin{equation}
    r^{\Zr\Ru}=\frac{\hat B}{\Delta Q}\,\frac{\Delta Z}{A},
    \label{eq:rubyRatio}
\end{equation}
which is equivalent to the definition used by STAR in Eq.~\eqref{eq:ratioSTAR}~\cite{STAR:2024lvy}. The two systems are thus described at the same temperature, volume, and net-baryon density, but with slightly different chemical potentials.

Note that the density-matched setup involves a choice of reference system. In the results shown below we use $\Zr{+}\Zr$ to define the reference net-baryon density: we first evaluate $n_B^{\Zr}$ at the nominal $\mu_B$ and then adjust $\mu_B$ in $\Ru{+}\Ru$ such that $n_B^{\Ru}=n_B^{\Zr}$ (at fixed $T$). One could equally well reverse this choice and match $n_B^{\Zr}$ to $n_B^{\Ru}$.  The resulting $r^{\Zr\Ru}$ differs only mildly between these two prescriptions, with the largest sensitivity appearing in peripheral bins. Physically, enforcing equal $n_B$ (together with $\mathrm{d}V/\mathrm{d}y$ matching) effectively induces a small remapping between the centrality percentiles of the two systems; because the neutron skin is larger in Zr than in Ru, this remapping is more pronounced for peripheral collisions, leading to a modest reference dependence that we treat as a controlled modeling systematic.

\paragraph*{(ii) Multiplicity-scaled setup (common $\mu_B$).}
As a second, more general prescription, we keep the same $\mu_B$ in both systems and introduce multiplicity scaling,
$\sigma^{\Ru}=\langle N^{\Ru}_{\rm ch}\rangle_{\rm ev}$ and
$\sigma^{\Zr}=\langle N^{\Zr}_{\rm ch}\rangle_{\rm ev}$,
evaluated for each centrality class. In this case the ratio is computed directly from Eq.~\eqref{eq:ratio}. Comparing setups (i) and (ii) provides a useful estimate of the sensitivity of the results to the treatment of system-size differences.

\begin{figure}
    \centering
    \includegraphics[width=0.48\textwidth]{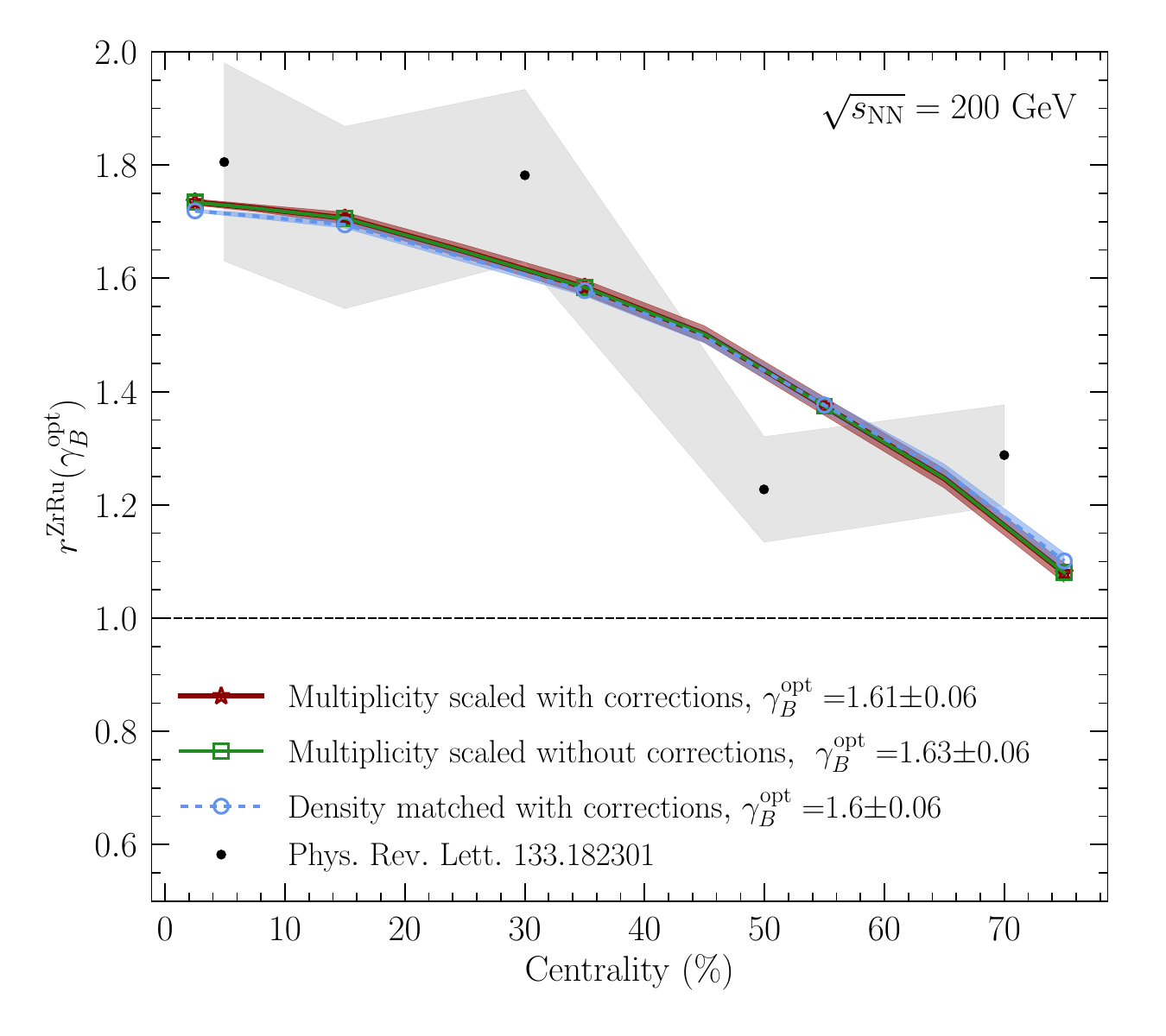}
    \caption{
    Centrality dependence of the isobar ratio $r^{\Zr \Ru}(\gamma_B^{\rm opt})$ at $\sqrt{s_{\rm NN}}=200$~GeV. The three curves correspond to the implementations indicated in the legend: (i) Multiplicity scaling with the corrections in Eq.~\eqref{eq:DeltaPhi}; (ii) Multiplicity scaling without corrections ($\epsilon^{\Ru}=\beta^{\Zr\Ru} = 0$); and (iii) the density matched $\mathrm{d}V^{\Ru}/\mathrm{d}y=\mathrm{d}V^{\Zr}/\mathrm{d}y$, and net-baryon density  $n_B^{\Ru}=n_B^{\Zr}$ by adjusting $\mu_B$. Black points (with the gray band) show the hydrodynamic proxy from Ref.~\cite{Pihan:2024lxw}. The dashed horizontal line at $r=1$ indicates the GCC expectation. The quoted values for $\gamma_B^{\rm opt}$ are the optimal values found from $\chi^2$ minimization in each case.
    }
    \label{fig:rISO}
\end{figure}

Figure~\ref{fig:rISO} shows our results. The three curves correspond to: (dashed lines, blue circles) the density-matched setup with $\sigma^{\Ru}=\sigma^{\Zr}=1$ and $\Delta\tilde Q$ computed including the correction factors $\tilde\beta^{XY}$ and $\tilde\epsilon^Y$ in Eq.~\eqref{eq:DeltaPhi}; (solid line, red stars) the multiplicity-scaled setup with $\sigma^{\Ru,\Zr}=\langle N_{\rm ch}^{\Ru,\Zr}\rangle_{\rm ev}$ and corrections factors; (solid line, green squares) the multiplicity-scaled setup with the corrections neglected, i.e.\ $\tilde\beta^{XY}=\tilde\epsilon^Y=0$. For all these cases, we extract the excess baryon-stopping parameter by a $\chi^2$ fit on the hydrodynamic simulations of Ref.~\cite{Pihan:2024lxw} used as a proxy for experimental data\footnote{Since the experimental $r$ data are presently preliminary~\cite{STAR:2024lvy} we use the simulated points of Ref.~\cite{Pihan:2024lxw} as it reproduces the qualitative trend of the preliminary STAR measurements, with minor quantitative differences.}

We find that all three implementations yield very similar results regarding the extracted baryon-stopping parameter (see the legend in Fig~\ref{fig:rISO}), all compatible within the fit uncertainty, as such, we consider in the remainder that 
\begin{equation}
    \gamma_B = 1.61\pm0.06
\end{equation}
In the same way, the centrality dependence of the density-matched and multiplicity scaled prescriptions are essentially the same. There is a very slightly weaker dependence of the density-matched which can be attributed to small differences in the overlap geometry and deformation effects between Ru and Zr which are washed out when the densities are matched. The $\tilde\beta^{XY}$ and $\tilde\epsilon^Y$ corrections are found to have a minor impact on $r^{\Zr\Ru}$, consistent with their small numerical size discussed in Appendix~\ref{sec:AppA} for the density-matched and the multiplicity scaled setups.
 
As expected from Sec.~\ref{sec:TheoreticalFramework}, the centrality dependence of $r^{\Zr\Ru}$ is driven predominantly by $p_{\rm frac}^X(c)$ (neutron-skin geometry), while varying $\gamma_B$ mainly produces an overall normalization shift with comparatively weak centrality dependence.

We also performed the estimation of $\gamma_B$ by calculating $\tilde{B}$ and $\tilde{Q}$ directly in \texttt{Thermal-FIST} rather than relying on proxies.
The results are $\gamma_B = 1.58\pm0.06$ for density-matched and $\gamma_B = 1.6\pm0.06$ for multiplicity scaled setups, which is compatible within error bars with the value obtained with the STAR proxy. 

We emphasize the importance of including the neutron skin effect.
In particular, our results show that taking the value of $r^{\Zr\Ru}$ in central collisions, where $\gamma_B \simeq 1.75$, as a naive estimate for extra baryon-stopping parameter $\gamma_B$ overestimates its actual value $\gamma_B \simeq 1.61$ by more than 10\%.
Although the effect of neutron skin of participant proton fraction may seem small in central collisions (Fig.~\ref{fig:pfracGlauber}), the net-charge difference $\Delta \tilde Q$ in the denominator of $r$ is extremely sensitive to it, leading to a sizable effect of the neutron skin even in central collisions.

These results demonstrate that the present framework can be used to extract $\gamma_B$ from precision measurements of the centrality dependence of $r$ in isobar collisions. 
More broadly, the approach provides a quantitative basis to study mechanisms that generate excess baryon stopping relative to charge stopping, such as baryon-junction transport, while controlling for nuclear-structure effects.
It also provides a quantitative cross-check of the experimentally necessary techniques developed and used by STAR~\cite{STAR:2024lvy}, namely the use of deuteron yields to estimate neutron contributions and the use of double ratios to determine the relative net-charge difference with reduced uncertainty.

\subsection{Neutron skin}

The centrality dependence of the generalized ratio $r^{\Gamma X}$ is, in general, controlled by the neutron-skin properties of \emph{both} nuclei $\Gamma$ and $X$. This entanglement can complicate the extraction of the neutron skin of a single nucleus from $r^{\Gamma X}$ measurements. A useful simplification arises when one of the two nuclei has a negligible neutron skin. If, for instance, nucleus $\Gamma$ has $\Delta R_{\rm np}\simeq 0$, then the centrality dependence of $r^{\Gamma X}$ predominantly reflects the neutron-skin effects in nucleus $X$.

Motivated by this observation, we propose to use $\Gamma = {}^{16}\mathrm{O}$---for which the neutron skin is negligible in our baseline description---as a reference system and to study oxygen-based ratios $r^{O X}$ in order to probe neutron skins of heavier nuclei $X$. Since the nuclei $X$ considered here (Table~\ref{table:WSParams}) differ substantially in size from oxygen, we employ multiplicity scaling,
$\sigma^{O}=\langle N^{O}_{\rm ch}\rangle_{\rm ev}$ and $\sigma^{X}=\langle N^{X}_{\rm ch}\rangle_{\rm ev}$, and evaluate $\Delta\tilde Q$ using the full expression in Eq.~\eqref{eq:DeltaPhi} (i.e.\ retaining the correction factors).

\begin{figure}
    \centering
    \includegraphics[width=0.45\textwidth]{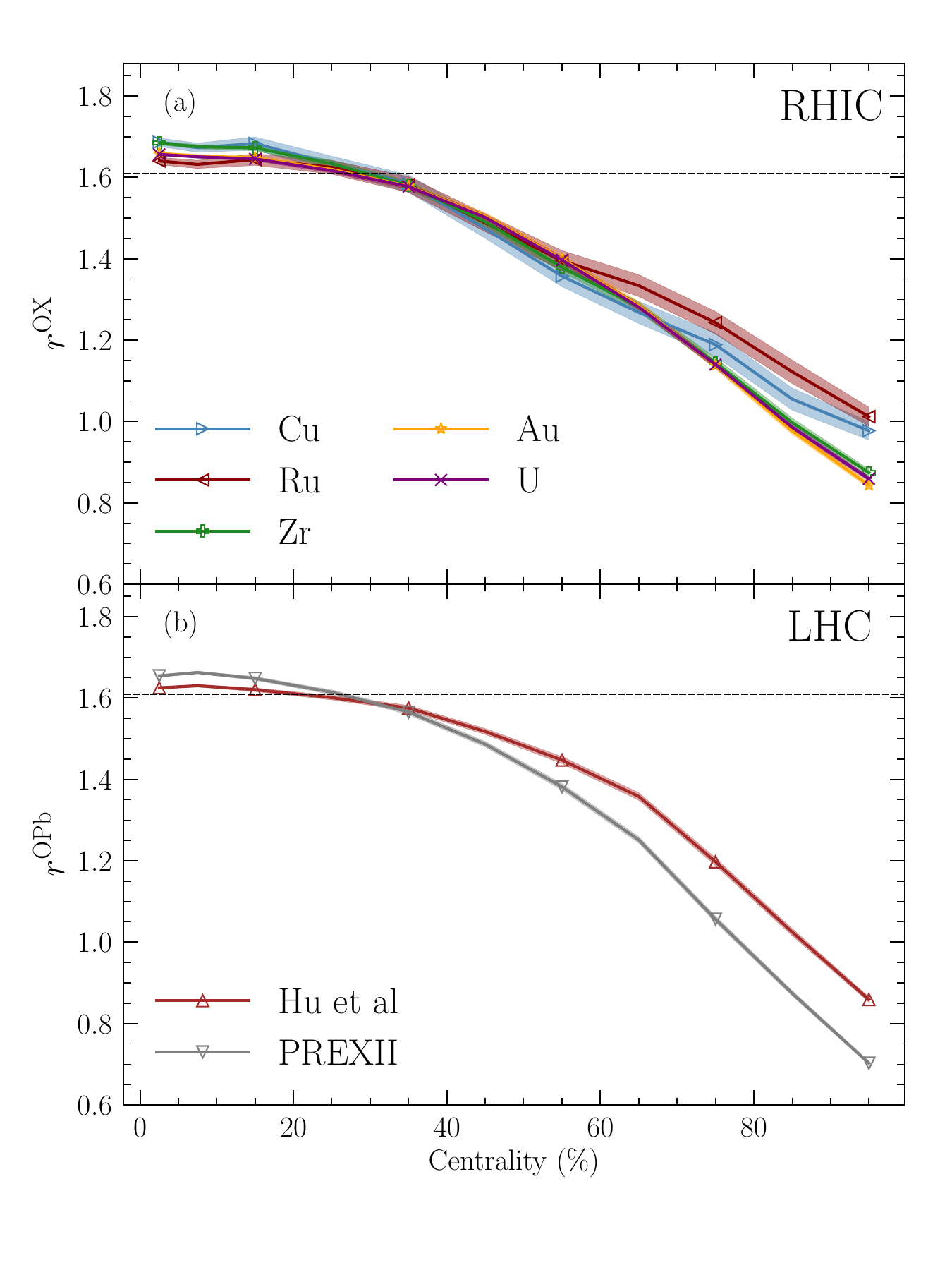}
    \caption{Centrality dependence of the oxygen-based ratio $r^{OX}$, evaluated using Eq.~\eqref{eq:ratio} with multiplicity scaling. Panel (a): $X=\Cu,\Ru,\Zr,\Au,\U$ at RHIC. Panel (b): $X=\Pb$ at LHC for two benchmark neutron-skin thicknesses, $\Delta R_{\rm np}\simeq 0.17$~fm and $\Delta R_{\rm np}\simeq 0.28$~fm (Table~\ref{table:WSParams}).
    }
    \label{fig:rOX}
\end{figure}

Figure~\ref{fig:rOX} shows the centrality dependence of $r^{OX}$ for the nuclei $X$ listed in Table~\ref{table:WSParams}, for RHIC [panel (a)] and LHC [panel (b)]. Throughout, we use $\gamma_B=1.61$, constrained from the isobar analysis in Sec.~\ref{sec:TheoreticalFramework}, including for LHC predictions.
We thus adopt the assumption that the relative excess of baryon over charge stopping, governed by the underlying QCD mechanism, varies slowly with collision energy, though a dedicated study of the energy dependence remains warranted.

In panel (a), the centrality dependence closely follows the behavior of the participant proton fraction in Fig.~\ref{fig:pfracGlauber}, underscoring the strong sensitivity of $r^{OX}$ to neutron-skin effects.
In panel (b), we show $r^{O\Pb}$ for two representative values of the Pb neutron-skin thickness. The overall amplitude of the ratio for LHC conditions is compatible with the ratio evaluated at RHIC energies. 
We observe that increasing $\Delta R_{\rm np}$ enhances the depletion of participant protons in peripheral Pb+Pb collisions and correspondingly increases the relative proton fraction in central collisions. This geometric effect translates into an increase of $r^{O\Pb}$ in central events and a decrease in peripheral events. These observations motivate introducing a robust neutron-skin proxy based on the \emph{central-to-peripheral} comparison,
\begin{equation}
\label{eq:ROX}
\mathcal{R}^{OX} \equiv \frac{r^{OX}(0\text{--}5\%)}{ r^{OX}(60\text{--}80\%)},
\end{equation}
which isolates the centrality-dependent component of the observable. Because $\gamma_B$ predominantly induces an overall normalization shift with relatively weak centrality dependence in our baseline implementation, the ratio $\mathcal{R}^{OX}$ is significantly less sensitive to $\gamma_B$ than the absolute magnitude of $r^{OX}$. 
We chose to use $0$--$5\%$ and $60$--$80\%$ centrality bins as they are typically accessible experimentally. Using more peripheral bins, for instance $80$--$100\%$ or $90$--$100\%$ would yield an even stronger impact from the neutron skin.

\begin{figure}
    \centering
    \includegraphics[width=0.45\textwidth]{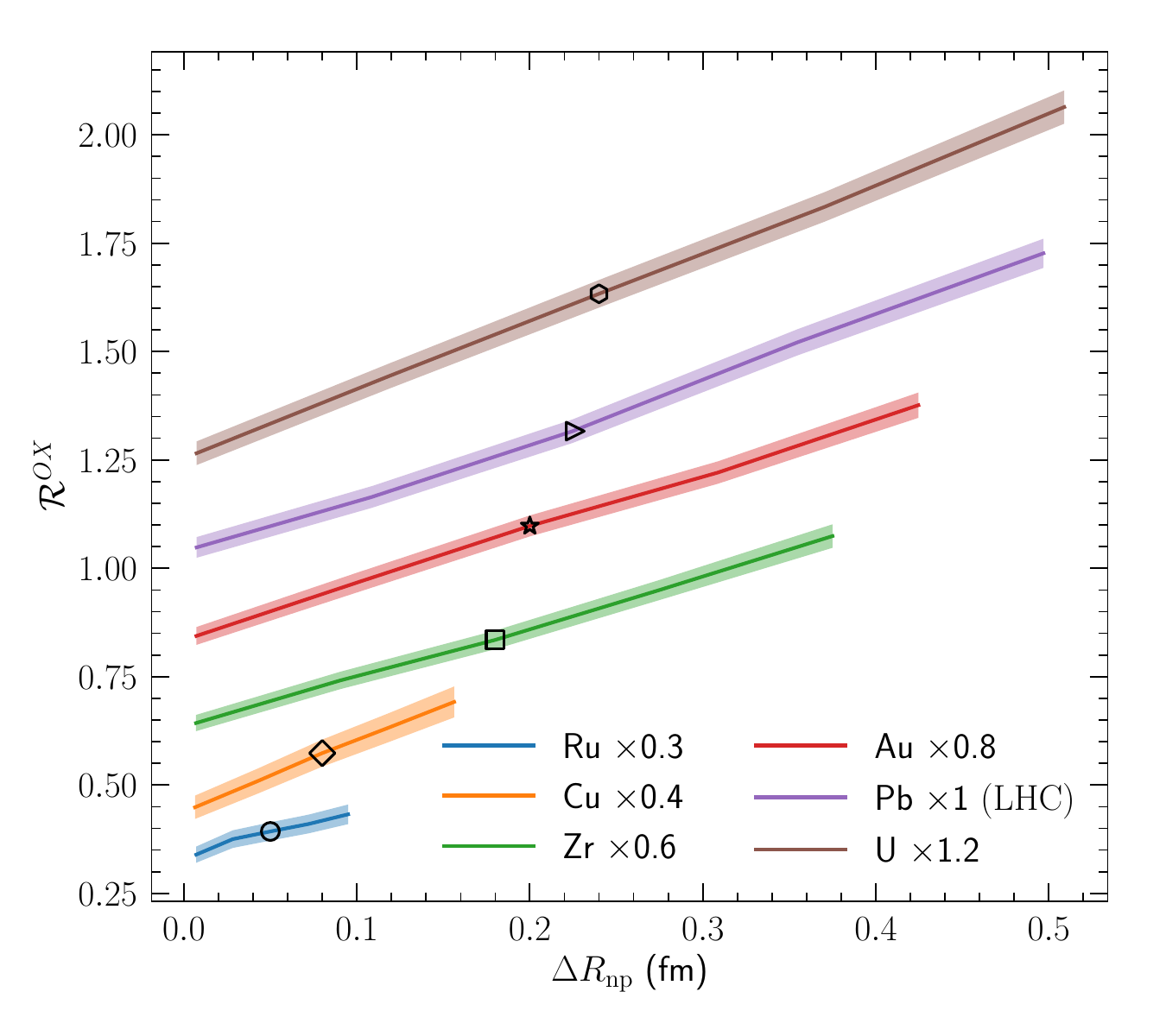}
    \caption{Central-to-peripheral oxygen-based ratio 
    ${\mathcal{R}^{OX}=r^{OX}(0\text{--}5\%)/r^{OX}(60\text{--}80\%)}$ as a function of neutron-skin thickness $\Delta R_{\rm np}$. Markers indicate literature values from Table~\ref{table:WSParams}. Multiplicative factors are applied for visual clarity.}
    \label{fig:DeltarOX}
\end{figure}

We therefore advocate $\mathcal{R}^{OX}$ as a particularly robust experimental handle on neutron-skin physics. To quantify its sensitivity, we vary the neutron skin of nucleus $X$ by adjusting the neutron diffuseness parameter $a_n$ in the Woods--Saxon profile while keeping the remaining parameters fixed.
Figure~\ref{fig:DeltarOX} shows the resulting dependence of $\mathcal{R}^{OX}$ on the RMS neutron-skin thickness $\Delta R_{\rm np}$ for different nuclei $X$ at both RHIC and LHC energies. 
An approximately linear dependence is observed across all systems considered, indicating that $\mathcal{R}^{OX}$ provides a sensitive handle on neutron-skin physics. 
We therefore advocate experimental measurements of oxygen-based central-to-peripheral differences, which can place stringent constraints on neutron-skin thicknesses for nuclei accessible in heavy-ion programs.

\section{Conclusion}

We developed a thermal-model framework to interpret midrapidity stopping observables that compare net baryon number and net electric charge in relativistic heavy-ion collisions. The key ingredients are (i) a centrality-dependent participant proton fraction $p_{\rm frac}^X(c)$ from the 3D Glauber model (encoding neutron-skin effects) and (ii) a phenomenological parameter $\gamma_B$ that quantifies excess baryon stopping relative to charge stopping, implemented via the constraint $n_Q/n_B=p_{\rm frac}^X(c)/\gamma_B$. This construction provides a controlled way to separate nuclear-structure effects from baryon-transport physics within a statistical model description of chemical freeze-out.

For isobar collisions (Ru+Ru and Zr+Zr at $\sqrt{s_{NN}}=200$ GeV), we showed that the ratio $r^{\Zr\Ru}$ can be used to extract $\gamma_B$ with good robustness against reasonable choices of normalization and against the small corrections entering the double-ratio construction of $\Delta Q$. Using the hydrodynamic results of Ref.~\cite{Pihan:2024lxw} as a proxy for the experimental trend, we obtained
$\gamma_B = 1.61\pm 0.06$.
This demonstrates that upcoming precision measurements of the centrality dependence of $r^{\Zr\Ru}$ in isobars can be converted into a quantitative determination of the extra baryon stopping component.

We then generalized the observable to asymmetric system pairs and proposed using oxygen as a baseline nucleus with negligible neutron skin. The resulting ratio $r^{OX}$ isolates the neutron-skin dependence of the target nucleus $X$ via centrality dependence, and the central-to-peripheral ratio
$\mathcal{R}^{OX}=r^{OX}[0\text{--}5\%]/r^{OX}[60\text{--}80\%]$ emerges as a particularly sensitive probe of $\Delta R_{np}$. 
Our results indicate that measurements of $\mathcal{R}^{OX}$ across available collision species could provide stringent constraints on neutron-skin thicknesses from heavy-ion data. 
A systematic study of the collision-energy dependence of $\gamma_B$ and of additional modeling systematics, such as oxygen clustering effects, choice of scaling parameters $\sigma^X$, or improved treatments of experimental proxies, is left for future work.

\section*{Acknowledgments}

We acknowledge useful discussions with Zhangbu Xu.
This work was supported by the U.S. Department of Energy, 
Office of Science, Office of Nuclear Physics, Early Career Research Program under Award Number DE-SC0026065.

\appendix

\section{Double ratio corrections\label{sec:AppA}}

In this Appendix we derive Eq.~\eqref{eq:DeltaPhi} and explicitly define the expansion parameters.
The derivation follows the strategy of Ref.~\cite{Lewis:2022arg}, adapted to our notations and to the
generalization from isobars to arbitrary system pairs.

The double ratio defined in Eq.~\eqref{eq:R2} can be related to the difference in net yields between two
collision systems $X$ and $Y$ only when (i) the charge asymmetry within system $Y$ and (ii) the mismatch between the yields in $X$ and $Y$ are both small. For isobar collisions this holds to very good accuracy, whereas for more disparate systems the deviations must be quantified. 
In addition, the choice of
scaling factors $\sigma^X$ and $\sigma^Y$ does not affect the double ratio itself, but it does modify the
overall magnitude of the net-yield differences appearing in Eq.~\eqref{eq:DeltaPhi}.

Starting from the definition
\begin{equation}
    R2^{XY}_{\tilde h} \equiv
    \frac{\tilde h_+^X/\tilde h_-^X}{\tilde h_+^Y/\tilde h_-^Y},
    \label{eq:AR2}
\end{equation}
we introduce the yield mismatches
\begin{align}
    \delta \tilde h^{XY}_+ &\equiv \tilde h_+^X - \tilde h_+^Y, \\
    \delta \tilde h^{XY}_- &\equiv \tilde h_-^X - \tilde h_-^Y .
    \label{eq:AR3}
\end{align}
Then
\begin{equation}
    R2^{XY}_{\tilde h} =
    \frac{1+\alpha}{1+\beta},
    \label{eq:AR4}
\end{equation}
with the (dimensionless) expansion parameters
\begin{align}
    \alpha &\equiv \frac{\delta \tilde h^{XY}_+}{\tilde h^Y_+}, \label{eq:Aalpha}\\
    \beta  &\equiv \frac{\delta \tilde h^{XY}_-}{\tilde h^Y_-}. \label{eq:Abeta}
\end{align}
These quantify the relative difference between systems $X$ and $Y$ for the positive- and negative-charge
yields of species $h$ (e.g.\ $h=\pi,K,p$) at midrapidity. They are not independent because $\tilde h^X_\pm$
and $\tilde h^Y_\pm$ are correlated through the $X$ and $Y$ mixing, but keeping both is convenient.

\subsection*{Charge asymmetry in the system $Y$}

We define the charge-averaged yield in $Y$,
\begin{equation}
    \hat h^Y \equiv \frac{\tilde h_+^Y+\tilde h_-^Y}{2},
\end{equation}
and the (half-)difference
\begin{equation}
    \delta \hat h^Y \equiv \frac{\tilde h_+^Y-\tilde h_-^Y}{2},
\end{equation}
so that $\tilde h^Y_\pm=\hat h^Y \pm \delta \hat h^Y$. The corresponding charge-asymmetry parameter is
\begin{equation}
    \tilde\epsilon^Y \equiv \frac{\delta \hat h^Y}{\hat h^Y}
    =\frac{\tilde h_+^Y-\tilde h_-^Y}{\tilde h_+^Y+\tilde h_-^Y}.
    \label{eq:AGamma}
\end{equation}
Importantly, $\tilde\epsilon^Y$ depends only on the intrinsic charge asymmetry of system $Y$ and is
independent of the overall scalings $\sigma^X$ and $\sigma^Y$ (whereas $\alpha$ and $\beta$ do depend on
the chosen scaling).

Using $\tilde h^Y_\pm=\hat h^Y(1\pm \tilde\epsilon^Y)$, we obtain
\begin{equation}
    \alpha-\beta
    =\frac{1}{\hat h^Y}\left[
    \frac{\delta\tilde h^{XY}_+}{1+\tilde\epsilon^Y}
    -\frac{\delta\tilde h^{XY}_-}{1-\tilde\epsilon^Y}
    \right].
    \label{eq:Aambeta}
\end{equation}
Assuming $|\tilde\epsilon^Y|<1$ (true at RHIC and LHC energies), we expand
\begin{equation}
    \frac{1}{1\pm \tilde\epsilon^Y}=\sum_{k=0}^{\infty} (\mp 1)^k (\tilde\epsilon^Y)^k,
\end{equation}
which yields
\begin{align}
    \alpha-\beta
    &=
    \frac{1}{\hat h^Y}\left[
    \sum_{k=0}^{\infty}
    \Big\{(-1)^k\,\delta\tilde h^{XY}_+ - \delta\tilde h^{XY}_-\Big\}
    (\tilde\epsilon^Y)^k
    \right]
    \nonumber\\
    &=
    \frac{1}{\hat h^Y}\left[
    (\delta\tilde h^{XY}_+ - \delta\tilde h^{XY}_-)\sum_{k\in 2\mathbb{N}_0}(\tilde\epsilon^Y)^k \right. \nonumber \\
    & \quad \left.
    -(\delta\tilde h^{XY}_+ + \delta\tilde h^{XY}_-)\sum_{k\in 2\mathbb{N}_0+1}(\tilde\epsilon^Y)^k
    \right].
\end{align}
Introducing the net-yield difference
\begin{equation}
    \Delta\tilde h \equiv \delta\tilde h^{XY}_+ - \delta\tilde h^{XY}_-,
\end{equation}
and noting that $\Delta\tilde h + (\delta\tilde h^{XY}_+ + \delta\tilde h^{XY}_-)=2\delta\tilde h^{XY}_+$,
we obtain the compact form
\begin{align}
    \alpha-\beta
    &=
    \frac{1}{\hat h^Y}\left[
    \frac{\Delta\tilde h}{1-\tilde\epsilon^Y}
    -2\,\delta\tilde h^{XY}_+\,\frac{\tilde\epsilon^Y}{1-(\tilde\epsilon^Y)^2}
    \right].
    \label{eq:Aalpha_minus_beta_final}
\end{align}

\subsection*{Final form and correction parameter $\theta$}
Using $R2^{XY}_{\tilde h}=(1+\alpha)/(1+\beta)=1+(\alpha-\beta)/(1+\beta)$, we find the double ratio
\begin{equation}
    R2^{XY}_{\tilde h}
    =
    1
    +
    \frac{\Delta\tilde h}{\hat h^Y(1+\beta)(1-\tilde\epsilon^Y)}
    \left(1-\theta\right),
    \label{eq:AR5}
\end{equation}
where the correction factor
\begin{equation}
    \theta
    \equiv
    2\,
    \frac{\delta\tilde h^{XY}_+}{\Delta\tilde h}\,
    \frac{\tilde\epsilon^Y}{1+\tilde\epsilon^Y}.
    \label{eq:Atheta}
\end{equation}
The parameter $\theta$ quantifies the deviation from the naive identification
$\Delta\tilde h \propto \hat h^Y (R2^{XY}_{\tilde h}-1)$, and becomes small when either the intrinsic charge
asymmetry in $Y$ is small or the system mismatch
is small.

Neglecting $\theta$ (and keeping the leading dependence on $\beta$ and $\tilde\epsilon^Y$) leads to
\begin{equation}
    \Delta \tilde h
    =
    (1+\tilde\beta^{XY})(1-\tilde\epsilon^Y)\,\hat h^Y\,
    \left(R2^{XY}_{\tilde h}-1\right),
    \label{eq:AR6}
\end{equation}
which is the form used in the main text [Eq.~\eqref{eq:DeltaPhi}].
Here we used $\tilde\beta^{XY}\equiv \beta$ to emphasize that $\beta$ depends on the chosen scalings
$\sigma^X$ and $\sigma^Y$.

\subsection*{Numerical size of $\theta$, $\epsilon^Y$ and $\beta^{XY}$ corrections}

\begin{table}[t]
\centering
\caption{Representative average values of $\theta$ [Eq.~\eqref{eq:Atheta}] for the isobar ratio
$r^{\mathrm{\Ru \Zr}}$. 
}
\begin{tabular}{|c|c|c|}
\hline
$| \langle\theta\rangle |$ & fixed volume & scaled \\
\hline
$\pi$ & $0.009$ & $0.0062$ \\
$p$   & $0.16956$   & $0.1142$ \\
$K$   & $0.02933$ & $ 0.01987$ \\
\hline
\end{tabular}
\label{table:ThetaIsobars}
\end{table}

\begin{table}[t]
\centering
\caption{Representative average values of $\theta$ [Eq.~\eqref{eq:Atheta}] for the oxygen-baseline ratio
$r^{\mathrm{O\,X}}$ at RHIC energies. 
Upper three rows correspond to calculation without scaling factors ($\sigma^X = \sigma^Y = 1$) while the bottom three rows incorporate multiplicity scaling factors, $\sigma^X=\langle N^X_{\rm{ch}}\rangle_{\rm{ev}}$, $\sigma^Y=\langle N^Y_{\rm{ch}}\rangle_{\rm{ev}}$.}
\begin{tabular}{|c|c|c|c|c|c|c|}
\hline
$| \langle\theta\rangle |$ & Cu & Ru & Zr & Au & U & Pb\\
\hline
$\pi$ & $0.9203$ & $0.93853$ & $0.89562$ & $0.92937$ & $0.92672$ & $0.93045$\\
$p$   & $1.00905$ & $1.00728$ & $1.01863$ & $1.01086$ & $1.01198$ & $1.01349$\\
$K$   & $1.02274$ & $1.01842$ & $1.0532$ & $1.02904$ & $1.03257$ & $1.03401$\\
\hline
$\tilde\pi$ & $0.00637$ & $0.0064$ & $0.00693$ & $0.00715$ & $0.00733$ & $0.00024$ \\
$\tilde p$  & $0.11462$   & $0.1147$   & $0.11349$   & $0.11316$   & $0.11284$ & $0.00373$\\
$\tilde K$  & $0.01987$& $0.01988$& $0.01949$& $0.01936$& $0.01924$ & $0.00063$\\
\hline
\end{tabular}
\label{table:ThetaOX}
\end{table}

Tables~\ref{table:ThetaIsobars} and~\ref{table:ThetaOX} summarize representative average values of $\theta$
for the ratios $r^{\mathrm{Zr\,Ru}}$ (isobars) and $r^{OX}$ (oxygen baseline), evaluated for two
choices of scaling: $\sigma^X=\sigma^Y=1$ (density matched) and $\sigma^X=\langle N^X_{\rm ch}\rangle_{\rm ev}$,
$\sigma^Y=\langle N^Y_{\rm ch}\rangle_{\rm ev}$ (multiplicity scaled). For the isobars, in both cases, we observe that the amplitude of $\theta$ is small compared to one for the pions and kaons but is sizable for the protons. This difference can be imputed to the baryon stopping, which increases the charge asymmetry as more protons than antiprotons are found at midrapidity. As we will see later (see Table~\ref{table:epsilonY}) the amplitude of the correction factor $\theta$ is largely driven by the charge asymmetry $\epsilon^Y$. Even though the correction is large for the protons, the net electric charge calculated using the double ratio is dominated by the pion contribution, as such neglecting $\theta$ for the protons remains a very good approximation in practice. 
For the $r^{OX}$ ratios the systems are much more different. Without scaling, $\theta$ is large, while multiplicity scaling reduces $\theta$ for $h=\pi,K$ to a negligible level. For protons
the residual correction can remain at the $\mathcal{O}(0.1)$ level, the scaling cannot compensate for the stopping. We emphasize that the double ratios cannot be used to evaluate the difference in net charge if the proton mismatch due to the baryon stopping becomes comparable to unity. In the same way as for the isobar case, in practice, $\Delta\tilde Q$ is dominated by the pion contribution, so the impact of the proton correction on the total net charge is small, which justifies neglecting $\theta$ also for the oxygen based calculations. We can thus express the net number $\Delta \tilde{h}$ as
\begin{equation}
    \Delta \tilde{h} = (1+\tilde{\beta}^{XY}) (1-\tilde{\epsilon}^Y) \hat{h}^Y(R2^{XY}_{\tilde{h}} - 1)
    \label{eq:AR6}
\end{equation}
The possible cancellation of the systematic errors using double ratios $R2_{\bar{h}}^{XY}$ ratio will be effective in $\Delta\tilde{h}$ only if $\beta^{XY}$ and $\epsilon^Y$, which explicitly breaks the cancellation, are negligible. In Tables~\ref{table:BetaXY} and~\ref{table:epsilonY} we show their numerical value for the isobars and the oxygen based ratios in the same cases as for the correction parameter $\theta$ (Tables \ref{table:ThetaIsobars}, \ref{table:ThetaOX}). The correction $\beta^{XY}$ is completely negligible for all cases except when the $\sigma^X=\sigma^Y=1$ in the oxygen based calculations. As $\beta^{XY}$ quantifies the mismatch between systems $X$ and $Y$, this observation follows naturally from the difference between $\Ox$ and the other systems. It shows that the multiplicity scaling is an excellent approach to control the systematic difference between different nuclei in the construction of the ratio $r^{OX}$. The correction $\epsilon^Y$ (for $Y=\Ru$ for the isobars) shows a very similar behavior as the correction parameter $\theta$ for all cases. This demonstrates that the corrections are largely driven by the charge asymmetry and in particular, the protons and antiprotons asymmetry. This mismatch is related to the baryon stopping showing that double ratios can, in principle, be used only when the proton yield due to the extra baryon stopping is small as compared to the total proton yield at midrapidity.

\begin{table}[t]
\centering
\caption{Representative average values of $\beta^{XY}$ [Eq.~\eqref{eq:beta}] for the isobars (top table) and the oxygen-baseline ratio
$r^{\mathrm{O\,X}}$ at RHIC energies (bottom table). 
In the bottom table, the upper three rows correspond to calculation without scaling factors ($\sigma^X = \sigma^Y = 1$) while the bottom three rows incorporate multiplicity scaling factors, $\sigma^X=\langle N^X_{\rm{ch}}\rangle_{\rm{ev}}$, $\sigma^Y=\langle N^Y_{\rm{ch}}\rangle_{\rm{ev}}$.}
\begin{tabular}{|c|c|c|}
\hline
$|\beta^{\Zr \Ru}|$ & Density-matched & Multiplicity scaling\\
\hline
$\pi$ & $0.00088$ & $0.00057$ \\
$p$   & $0.00066$ & $0.00094$ \\
$K$   & $0.00039$ & $0.00035$  \\
\hline
\end{tabular}
\begin{tabular}{|c|c|c|c|c|c|c|}
\hline
$|\beta^{\Ox X}|$ & Cu & Ru & Zr & Au & U & Pb\\
\hline
$\pi$ & $0.61129$ & $0.71833$ & $0.70059$ & $0.80575$ & $0.81434$ &  $0.82012$ \\
$p$  & $0.61143$   & $0.71844$   & $0.70084$   & $0.80596$   & $0.81457$ & $0.82012$ \\
$K$  & $0.61121$& $0.71827$& $0.70044$& $0.80563$& $0.81421$ & $0.82011$\\
\hline
$\tilde \pi$ & $0.00049$ & $0.00051$ & $0.00105$ & $0.00126$ & $0.00144$ & $5.10^{-5}$ \\
$\tilde p$  & $0.00082$   & $0.00084$   & $0.00176$   & $0.0021$   & $0.0024$ &  $8.10^{-5}$\\
$\tilde K$  & $0.00031$& $0.00032$& $0.00063$& $0.00079$& $0.00093$ & $3.10^{-5}$ \\
\hline
\end{tabular}
\label{table:BetaXY}
\end{table}

\begin{table}[t]
\centering
\caption{Representative average values of $\epsilon^{Y}$ [Eq.~\eqref{eq:epsilon}] for the isobars (top table) and the oxygen-baseline ratio
The correction $\tilde{\epsilon}^Y$ does not depend on the choice of $\sigma^X$ and $\sigma^Y$, the bottom panel thus only represents the multiplicity scaled, $\sigma^X=\langle N^X_{\rm{ch}}\rangle_{\rm{ev}}$, $\sigma^Y=\langle N^Y_{\rm{ch}}\rangle_{\rm{ev}}$ values.}
\begin{tabular}{|c|c|c|}
\hline
$|\epsilon^{\Ru}|$ & Density matched & Multiplicity scaling \\
\hline
$\pi$ & $0.00629$ & $0.0063$ \\
$p$   & $0.11334$ & $0.11383$ \\
$K$   & $0.01965$ & $0.01974$  \\
\hline
\end{tabular}
\begin{tabular}{|c|c|c|c|c|c|c|}
\hline
$|\epsilon^{Y}|$ & Cu & Ru & Zr & Au & U & Pb \\
\hline
$\tilde \pi$ & $0.00628$ & $0.0063$ & $0.00685$ & $0.00706$ & $0.00724$ & $0.00024$ \\
$\tilde p$  & $0.11386$   & $ 0.11383$   & $0.11293$   & $0.11258$   & $0.11229$ & $0.00373$  \\
$\tilde K$  & $0.01975$& $0.01974$& $0.0194$& $0.01927$& $0.01916$ &  $0.00063$\\
\hline
\end{tabular}
\label{table:epsilonY}
\end{table}

\bibliography{biblio}

\end{document}